[1994/12/01]
\documentclass[10pt,a4paper]{article}
\usepackage{graphicx}
\title{\bf A note on the violation of Bell's inequality}
\author{Thomas Sch\"urmann\\
D\"usseldorf, Germany\\
\\
}

\usepackage{amsmath,amsthm}
\begin{document}
\maketitle
\renewcommand{\sectionmark}[1]{}
\vspace{.1in}
{\bf Abstract:}$\;\;$
With Bell's inequalities one has a formal expression to show how essentially all local theories of natural phenomena that are formulated within the framework of realism may be tested using a simple experimental arrangement. For the case of entangled pairs of spin-1/2 particles we propose an alternative measurement setup which is consistent to the necessary assumptions corresponding to the derivation of the Bell inequalities. We find that the Bell inequalities are never violated with respect to our suggested measurement process.\\
\\
{\bf Keywords:} Bell inequalities; EPR-Paradoxon; Quantum mechanics\\
\\

The argument of Einstein, Podolsky, and Rosen (EPR) \cite{EPR}, in 1935, concerning the completeness of quantum mechanics and the possible existence of hidden variables was originally couched in terms of position and momentum coordinates of a pair of particles which could assume a continuous range of values. Subsequently, Bohm \cite{Bohm51} put the argument in terms of an initially spin-0 system which dissociates into two spin-1/2 systems, the components of the spin of which could only take on discrete values. The critical questions raised by these authors inspired many researchers to study the essential difference between quantum physics and the classical concepts of reality and locality. The breakthrough was the discovery by Bell \cite{Bell64} that the statistical correlations of entangled quantum states are incompatible with the predictions of any theory which is based on concepts of reality and locality of EPR. The constraints imposed on statistical correlations within the framework can be cast into the form of inequalities, which are now generally referred to as Bell inequalities, which allow a quantitative distinction to be made between the prediction of quantum mechanics an local realism. \\

Let us briefly review the main idea of Bell. Suppose a pair of spin-1/2 particles which have been prepared somehow in a singlet state such that they move in different directions towards two measurement devices, and that these devices measure spin components along directions $\hat{a}$ and $\hat{b}$ respectively. Suppose that the hypothetical complete description of the initial state is in terms of hidden variables $\lambda$ with normed probability distribution $\rho(\lambda)$ for the given quantum-mechanical state. The result $A(=\pm 1)$ of the first measurement can clearly depend on $\lambda$ and on the setting $\hat{a}$ of the first instrument. Similar, $B$ can depend on $\lambda$ and $\hat{b}$. But the notion of locality requires that $A$ does not depend on $\hat b$, nor $B$ on $\hat a$. The results of the two selections are then to be deterministic functions with $A(\hat{a},\lambda)=\pm 1$ and $B(\hat{b},\lambda)=\pm 1$. In this terminology one can ask, if the mean value $P(\hat{a},\hat{b})$ of the product $AB$, i.e.  \\
\begin{equation}\label{Pab}
P(\hat{a},\hat{b})=\int d\lambda\;\rho(\lambda)\,
A(\hat{a},\lambda)\,B(\hat{b},\lambda)
\end{equation}
can equal the quantum-mechanical prediction. \\
Now, let $\hat{a}'$ and $\hat{b}'$ be alternative settings of the measurement instruments. Then, by simple algebraic manipulations one obtains \cite{CHHS69}
\begin{equation}\label{CHHS}
|P(\hat{ a},\hat{ b})-P(\hat{ a},\hat{ b}')|+P(\hat{ a}',\hat{ b}')+ P(\hat{a}',\hat{ b})\leq 2. \\
\end{equation}
If we consider the case in which the total spin is zero then we have
\begin{equation}\label{P1}
P(\hat{ b}',\hat{ b}')=-1.
\end{equation}
This is predicted by quantum mechanics and by any theory that defines the angular momenta of composite sytems. Then, for the special choice
$\hat{a}'=\hat{b}'$ equation (\ref{CHHS}) yields
\begin{equation}\label{Bell}
|P(\hat{a},\hat{b})- P(\hat{a},\hat{b}')|\leq 1+P(\hat{b}',\hat{b}).\\
\end{equation}
Inequalities (\ref{Bell}) constitute the original form of the Bell inequalities \cite{Bell64}. The generalized form (\ref{CHHS}) was first derived by Clauser, Horne, Shimony and Holt (CHSH)\cite{CHHS69}. Several researchers have since formulated further versions of the Bell inequalities e.g. \cite{Wigner70} \cite{BCH74} \cite{CS78} that are more experimentally amenable.\\

To study whether these inequalities hold for any unit vectors ${\hat a}$, ${\hat b}$, ${\hat b}'$, let the particles be initially prepared in a singlet state
\begin{equation}\label{singlet}
|\phi\rangle=\frac{1}{\sqrt{2}}\;(|+z,-z\rangle-|-z,+z\rangle)
\end{equation}
in $z$-direction. Then $P(\hat{a},\hat{b})$ is given by the corresponding quantum mechanical expectation value
\begin{equation}\label{QM}
\langle {\bf\sigma_1}\hat{a}\otimes{\bf\sigma_2}\hat{b} \rangle
= -\hat{a}\cdot\hat{b}.
\end{equation}
This function has the property (\ref{P1}). A simple instance of disagreement between the predictions of (\ref{QM}) and (\ref{Bell}) is provided by taking $\hat{ a}$, $\hat{ b}$ and $\hat{ b}'$ to be coplanar, with $\hat{b}'$ making an angle of $2\pi/3$ with $\hat{a}$, and $\hat{b}$ making an angle of $\pi/3$ with both $\hat{a}$ and $\hat{b}'$. Then $\hat{a}\cdot\hat{b}=\hat{b}\cdot\hat{b}'=1/2$ and $\hat{a}\cdot\hat{b}'=-1/2$. These values do not satisfy (\ref{Bell}). Hence the quantum-mechanical prediction, based on the expression (\ref{QM}), and that of (\ref{Bell}) are incompatible, at least for some pairs of analyser orientations.\\
\\
At this point we want to mention that Bell's violation argument is based on the particular choice of the quantum-mechanical measurement process corresponding to (\ref{QM}), for which the condition
\begin{equation}
\langle {\bf\sigma_1}\hat{a}\otimes{\bf\sigma_2}\hat{b}
-{\bf\sigma_1}\hat{a}\otimes{\bf\sigma_2}\hat{b}'\rangle
= \langle {\bf\sigma_1}\hat{a}\otimes{\bf\sigma_2}\hat{b}\rangle
-\langle {\bf\sigma_1}\hat{a}\otimes{\bf\sigma_2}\hat{b}'\rangle,
\end{equation}
is trivially satisfied. Thus, in Bell's approach, any experimental determination of the two correlation averages $\langle{\bf\sigma_1}\hat{a}\otimes{\bf\sigma_2}\hat{ b} \rangle$ and $\langle {\bf\sigma_1}\hat{a}\otimes{\bf\sigma_2}\hat{ b}' \rangle$ is based on (at least) two different sets of particle pairs, corresponding to the left hand side of (\ref{Bell}), and (at least) four different sets of particle pairs in the case of (\ref{CHHS}). \\
On the other hand, Bell's derivation of inequality (\ref{Bell}) is explicitly based on the assumtion that the values of $A(\hat{ a},\lambda)$, $B(\hat{ b},\lambda)$ and $B(\hat{ b}',\lambda)$ are considered for any individual pair of particles separately \cite{Bell64}, and without this assumption the inequality could not be derived. The latter the main reasons why we are not convinced to apply the measurement process suggested by Bell (and CHSH). Instead we suggest to apply a measurement process whose expectation value, is not necessarily measured on different sets of particles. The differnce of the measurement results between the conventional approach suggested by Bell and CHSH to our suggestion becomes obvious, because the commutator relation $[{\bf\sigma}\hat{b},{\bf\sigma}\hat{ b}']=i2{\bf\sigma}(\hat{b}\times\hat{ b}')$ does not vanish for all analyser orientations $\hat{b}$ and $\hat{ b}'$. \\
Let us describe the measurement processes in more detail. Without loss of generality we assume that the particle paire is initially prepared in $\hat{a}$-direction, i.e.
\begin{equation}
|\phi\rangle=\frac{1}{\sqrt{2}}(|+a,-a\rangle\;+\;|-a,+a\rangle)
\end{equation}
This state is then reduced by the first analyser with orientation in the direction of $\hat{a}$. In this setup, with probability $p_\pm(a)=1/2$, one finds the pair of particles in one of the reduced states $|\pm a,\mp a\rangle$. Subsequently it follows the next measurement by the second analyser with orientation in the direction of $\hat{b}$. The joint probabilities of both reductions are simply
\begin{equation}
p_{s_1s_2}(\hat{a},\hat{b})=\frac{1}{4} (1-s_1s_2\cos\Theta_{ab})
\end{equation}
with $\Theta_{ab}$ defined as the angle between $\hat{a}$ and $\hat{b}$ and $s_1,s_2=\pm 1$.
The final measurement is once more at the second particle, but the orientation of the analyser is $\hat{b}'$. Computation of the corresponding probability amplitues of the final state reduction yields the following normed joint probabilities for the total measurement process
\begin{equation}\label{meas}
p_{s_1s_2{s'_2}}(\hat{a},\hat{b},\hat{b}') =\frac{1}{8}(1-s_1s_2\cos\Theta_{ab})(1-s_2s'_2\cos\Theta_{bb'}),
\end{equation}
with $s_1,s_2,{s'_2}=\pm 1$. This eight probabilities represent our quantum-mechanical measurement process with respect to the assumptions in Bell's derivation of his inequalitiy (\ref{Bell}). Applying this probabilites to compute the correlation functions in the left hand side of (\ref{Bell}), we find
\begin{equation}\label{exp0}
|\langle{\bf\sigma_1}\hat{a}\otimes({\bf\sigma_2}\hat{b}-{\bf\sigma_2}\hat{b}')\rangle_0| =
 (1-\cos\Theta_{bb'})\;|\cos\Theta_{ab}|,
\end{equation}
where $\langle\;\rangle_0$ denotes the mean value corresponding to the joint probabilities (\ref{meas}). In contrast, the corresponding expression of the measurement process suggested by Bell is
\begin{equation}\label{exp1}
|\langle{\mathbf\sigma_1}\hat{a}\otimes({\bf\sigma_2}\hat{ b}-{\bf\sigma_2}\hat{ b}')\rangle| =|\cos\Theta_{ab'}-\cos\Theta_{ab}|.
\end{equation}
At this point, the difference between Bell's argument (\ref{exp1}) and our view concerning the measurment process (\ref{exp0}) becomes formal. \\
Much simpler is the situation corresponding to the right hand side in (\ref{Bell}). In this case we agree to measure the correlation function on different set of particle pairs because there is no contradiction to Bell's derivation of (\ref{Bell}) and thus follow his argumentation to compute the correlation function directly by (\ref{QM}), and get the ordinary result $-\cos\Theta_{b'b}$. \\
As we know, Bell's inequality is violated by the measurement process corresponding to expression (\ref{exp1}).
However, the situation is quite diffenerent when applying expression (\ref{exp0}) to the left hand side in (\ref{Bell}). As can be simply verified
a violation of Bell's inequality is not possible for any orientations $\Theta_{b'b}$ or $\Theta_{ab}$ of the analyzers, because the following expression holds
\begin{equation}
(1-\cos\Theta_{b'b})(1-|\cos\Theta_{ab}|)\geq 0.
\end{equation}
Therefore, Bell's inequality (\ref{Bell}) is never violated with respect to the quantum-mechanical measurement process corresponding to (\ref{meas}).
%
\newpage
%

\end{document}